# Use of Feedback to Maximize Photon Count Rate in XRF Spectroscopy

Benjamin A Lucas

**Abstract:** The effective bandwidth of an energy dispersive x-ray fluorescence spectroscopy system is limited by the timing of incident photons. When multiple photons strike the detector within the processing time of the detector photon pile-up occurs and the signal received by the detector during this interval must be ignored. In conventional ED-XRF systems the probability of a photon being incident upon the detector is uniform over time, and thus pile-up follows Poisson statistics. In this paper we present a mathematical treatment of the relationship between photon timing statistics and the count rate of an XRF system. We show that it is possible to increase the maximum count rates by applying feedback from the detector to the x-ray source to alter the timing statistics of photon emission. Monte-Carlo simulations, show that this technique can increase the maximum count rate of an XRF spectroscopy system by a factor of 2.94 under certain circumstances.

**Introduction:** Energy dispersive X-ray fluorescence spectroscopy (ED-XRF) is a powerful technique for the rapid non-destructive analysis of materials[1] that has found applications in fields ranging from material science to archeology [2],[3],[4]. In general, an energy dispersive X-ray fluorescence system consists of an x-ray tube along with a solid-state x-ray spectrometer (see figure 1). The x-ray source emits x-rays which excite inner shell electrons of the atoms in a test sample. This excitation results in the emission of the characteristic K and L lines of the elements present in the sample, and the intensity of these characteristic lines are directly related to the molar composition of the analyte material [5].

Since the sample's elemental composition is calculated from the intensity of the observed spectral lines, the accuracy of this analysis follows from the accuracy of the measurement of the intensity of the observed fluorescent peaks in the sample spectrum [6]. Since the photons are emitted randomly from the sample, the measurement of the intensity of a spectral line follows a Doisson distribution [6]. Thus the accuracy of the measurement of a line's intensity is given as $\sigma = 1/\sqrt{n}$, where $n$ is the number of photon counts in the peak. From this result, it follows that the accuracy of XRF analysis is determined by the photon count rate on the detector and the test duration. For many applications, time is a limiting factor, so the key optimization parameter is photon count rate.

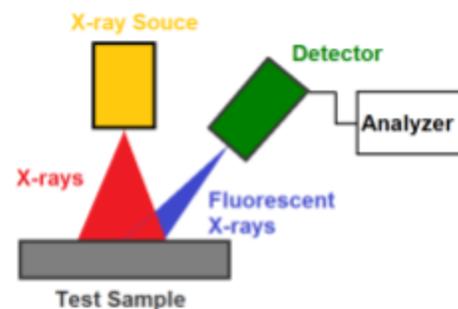

Figure 1.) Schematic of the basic components of an ED-XRF System

The photon count rate is limited by the speed at which photons can be detected and processed. Since photons are emitted from the analyte at random times, some portion of photons will arrive at the detector while it is still processing previous photons - an effect known as photon pileup. When pileup occurs, the energies of the constituent photons cannot be accurately distinguished, and the data from this time interval must be discarded. This effectively shrinks the proportion of time when the detector is operational, and can result in a significant reduction in the count rate of an XRF system. As a consequence of this most ED-XRF systems operate at an effective count rate that is less than 38% of the photon processing rate of the detector's electronics. In the paper we review the theory of underlying the timing statistics of ED-XRF systems and present a theoretical outline of how it might be possible to increase the effective bandwidth of an XRF system by altering these statistics.

**Theory:**
The relationship between the the incident count rate (ICR) and the output count rate (OCR) is a simple matter of the proportion of incident counts that do not result in pileup.

$$OCR = P_{good} \cdot ICR \qquad \text{Eq 1}$$

For an XRF spectroscopy system, the theoretical upper bound on the count rate is given by a system which does not have any pile-up events, and is thus solely by the speed of the detector's photon processing. If a detector can process a single photon in $\mu$ seconds, then the maximum number of photons that can be processed during a one second time interval is given as the $OCR_{max} = 1/\mu$ (Eq 2). For most solid-state detectors, $\mu$ ranges between 2us to 20 us giving an upper limit on the detector throughput ranging between 5e4 to 5e5 counts per second. This upper bound assumes that photons are detected at regular time intervals with exactly one photon being detected every $\mu$ seconds. In reality, photons are detected at random times with a probability *p* that is uniform across time. Since the probability of a photon being detected is constant, the number of photons detected at a given time follows the Poisson Distribution. The probability that given photon that incident upon the detector will not result in pile-up is the probability that there are no photons incident upon the detector for $\mu$ seconds before and $\mu$ seconds after the photon was detected. For a Poisson process, the probability that no event occurs is given as:

$$P(0) = e^{-\lambda} \qquad \text{Eq 3}$$

Where $\lambda$ is the average rate at which events take place. Since the detector takes $\mu$ seconds to process a photon, $\lambda = \mu \cdot ICR$, and $P(0) = e^{-\mu \cdot ICR}$.

Since a photon does not result in pile-up only if there are no counts before and after its time of arrival, the probability of a photon being processed without pileup is:

$$P_{good} = P(0)_{before} \cdot P(0)_{after} \qquad \text{Eq 4}$$

$$= (e^{-\mu \cdot ICR}) \cdot (e^{-\mu \cdot ICR}) \qquad \text{Eq 5}$$

$$= e^{-2\mu \cdot ICR} \qquad \text{Eq 6}$$

From equations 1 and 6, we see that the output count rate of a standard XRF system is

$$OCR = ICR \cdot e^{-2\mu \cdot ICR} \qquad \text{Eq 7}$$

By differentiating this result, we find that the maximum count rate occurs at $ICR_{max} = 1/(2\mu)$, and thus $OCR_{max} = 1/(5.42 \cdot \mu)$ (Eq 8).

The preceding treatment of a conventional XRF system shows the mathematical limitations of this design. If we compare equation 8 with the theoretical maximum in equation 2, we see that the maximum count rate is 5.42 times less than the limit allowed by the detector electronics alone. These limitations follow directly from the fact that the probability of photon emission is statistically uncorrelated with the detection of photons are previous times. Whenever this is the case, the output count rate will be less than or equal to the relationship outlined in eq 7. This result could change, however, if the probability of photon emission were modulated by the detection of photons. Hence it is possible achieve a higher maximum count rate on an XRF system by altering the photon timing statistics.

**Computational Modeling:**
When the probability of photon emission is effected by the detection of a photon, it becomes difficult to describe with a simple mathematical relationship. To overcome this limitation it is useful to employ computational methods to model the throughput of the xrf system. A monte carlo model of an XRF system was created in Python. The incidence of photons on the detector is represented by some probability distribution. The incidence of photons on the detector was then modeled by comparing this probability function to the output of a random number generator. The output count rate was then determined by finding the number of events which were detected which did not overlap with the the pulse processing time of other events.

As a test case, this computational model was run on a uniform emission probability, which mirrors the mathematical model outlined by equation 7. Our model assumed a pulse procession time of $\mu = 10\ us$, which according to equation 8 gives a peak count rate of 1.84e4 counts per second. The result of the this simulation along with the theoretical result obtained by equation 7 is shown in figure 2.

It can be seen that the computational model and the mathematical model from the previous section show good agreement

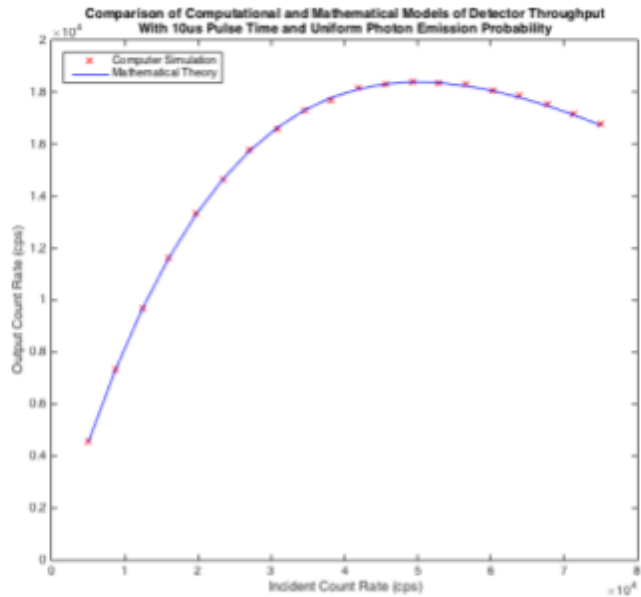

Figure 2.) Comparison the computational and mathematical relationship of the detector throughput for a uniform photon detection probability.

($R^2 > 0.99$), and the peak output count rate fall at the expected value predicted by equation 8.

Following up on this result, we model the throughput of an XRF system with simple feedback mechanism. Here the feedback mechanism is modeled such that the probability of photon emission is constant until a the detection of an event. Following this, the probability is held constant for a delay time of $\tau = 1\ us$ and then the probability is set to 0 for the subsequent $\mu - \tau$ seconds. For consistency, we have used a pulse time of $\mu = 10\ us$. The throughput of this system is plotted in comparison to the throughput of a system with a uniform detection probability in figure 3.

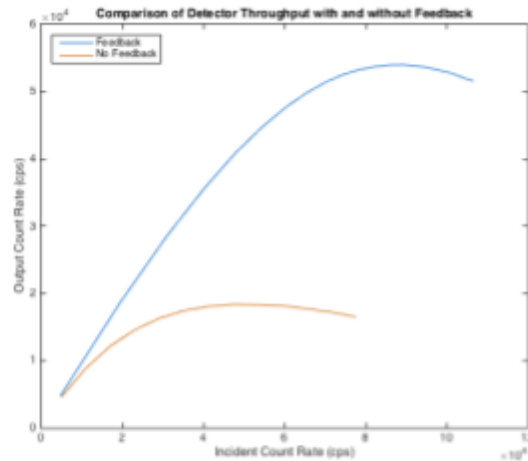

As can be seen, the computational model of an xrf system with feedback shows a marked gain in performance over the system with a uniform detection probability. The system with feedback achieves a maximum output count rate of 5.39e4, which constitutes a gain of 2.94 over a system with no feedback.

Figure 3.) Comparison of computational models of the detector throughput relationship a system with feedback to one without feedback to the source.

**Discussion:**
As the preceding mathematical and computational analysis indicates, it is theoretically possible to increase the efficiency of an ED-XRF system beyond the limit of Poisson statistics by modulating the photon timing statistics. The preceding analysis was done by with a feedback mechanism that had a delay time that was 10% of of the detector pulse time. As a practical concern, a faster feedback mechanism will be necessary for shorter pulse times. One potential implementation of feedback from the detector to the source would be to apply the detector signal as the input trigger of a monostable oscillator connected to the grid potential of the x-ray tube. Since the photon detection probability is directly proportional to the current [7] of the x-ray tube, changing the grid potential has a direct effect on the detection probability.

**Conclusions:**
The preceding work shows that the maximum count rate that can be achieved by an XRF system is limited not only by the processing speed of the detector, but by the timing of photons on the detector. For any XRF system where the probability of emission and probability of photon detection are uncorrelated, the count rate, and thus the accuracy of analysis is limited by Poisson statistics. Computational modelling indicates that a simple feedback mechanism between the detector and source, could possibly raise the count rate of an XRF system by a factor of 2.94, which constitutes a significant gain in performance. This result suggests a promising direction for further developments in XRF.

**References:**
[1] Jenkins, Ron. "X‐Ray Techniques: Overview." *Encyclopedia of analytical chemistry* (2000).

[2] Hein, M., et al. "Application of X-ray fluorescence analysis with total-reflection (TXRF) in material science." *Fresenius' journal of analytical chemistry* 343.9-10 (1992): 760-764.

[3] Moioli, Pietro, and Claudio Seccaroni. "Analysis of art objects using a portable X‐ray fluorescence spectrometer." *X Ray Spectrometry* 29.1 (2000): 48-52.

[4] Shackley, M. Steven. *An introduction to X-ray fluorescence (XRF) analysis in archaeology*. Springer New York, 2011.

[5] Sherman, J. "The Theoretical Derivation of Fluorescent X-ray Intensities from Mixtures." Spectrochimic. Acta, 7, (1955): pp. 283.

[6] Jenkins, Ron. *Quantitative X-ray spectrometry*. CRC Press, 1995.

[7] Huda, Walter. "Chapter 1." *Review of Radiologic Physics*. Baltimore, MD: Lippincott Williams & Wilkins, 2010.